\documentclass[prl,twocolumn]{revtex4-1}
\usepackage{amsmath,amssymb,bm}
\usepackage{graphicx}
\usepackage{epstopdf}
\usepackage{latexsym}
\usepackage{subfigure}
\usepackage{color}
\usepackage{times}
\usepackage{hyperref}

\begin{document}

\title{Quantum phase transitions in bilayer SU($N$) anti-ferromagnets}
\author{Ribhu K. Kaul}
\affiliation{Department of Physics \& Astronomy, University of Kentucky, Lexington, KY-40506-0055}
\begin{abstract}
We present a detailed study of the destruction of 
SU($N$) magnetic order in square lattice bilayer anti-ferromagnets
using unbiased quantum Monte Carlo numerical simulations and field theoretic
techniques. We study phase transitions from an SU($N$) N\'eel state into
two distinct quantum disordered ``valence-bond'' phases: a
valence-bond liquid (VBL) with no broken
symmetries
and a lattice-symmetry breaking valence-bond solid (VBS)
state. For finite inter-layer coupling, the cancellation of Berry
phases between the layers has dramatic consequences on the two phase
transitions: the N\'eel-VBS transition is first order for all $N\geq5$
accesible in our model, whereas
the N\'eel-VBL transition is continuous for $N=2$ and first order for
$N\geq 4$; for $N=3$ the N\'eel-VBL transition show no signs of
first-order behavior.
\end{abstract}
\maketitle

The
study of quantum phase transitions is an exciting field at the
forefront of theoretical condensed matter
physics~\cite{sachdev1999:qpt}.
The nature of a particular quantum phase transition is governed by
properties that affect long distance physics such as broken
symmetries, topological order, the presence
of Berry phases and is generally insensitive to microscopic details. 
Quantum magnets provide the richest examples of quantum phase
 transitions because they possess internal symmetries in
 addition to the usual lattice and time reversal symmetries and because
 they often have non-trivial Berry phases in their long wavelength descriptions~\cite{haldane1988:berry}.
The most popular internal symmetry group in condensed matter is the
SU($N$) group. Initial interest was focussed on SU($2$) and the case of $N>2$ was
introduced purely as a theoretical tool to access the
analytically solvable $N\rightarrow \infty$ limit~\cite{affleck1985:lgN,read1990:vbs}.  
However, in the ensuing years it has come to be recognized that
SU($N$) systems
with $N>2$ {\em but
finite} are interesting in their own right, since they serve to
model a number of physical systems ranging from spin-orbit coupled solid-state materials~\cite{kugel1982:kk} to ultra-cold atoms in optical lattice
potentials~\cite{gorshkov2010:sun}. While the ground states of SU($N$) spin models in
one-dimensional chains are relatively well understood~\cite{affleck1987:critchain,fuhringer2008:sun},
two-dimensional phases~\cite{hermele2009:sun,toth2010:su3,corboz2011:su4} and their associated phase
transitions are only poorly understood.

\begin{figure}[t]
\includegraphics[width=3.3in, clip=true, trim= 0 0 0 100]{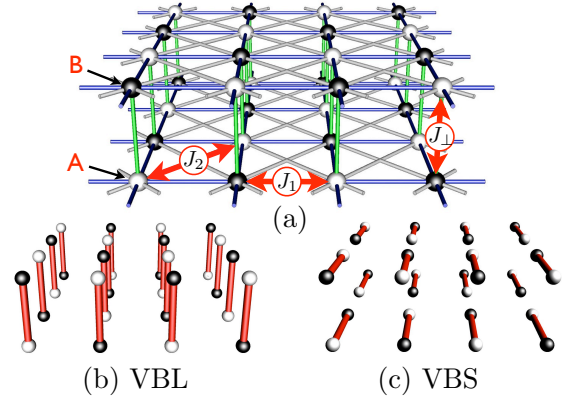}
\caption{\label{fig:intro} (a) Bilayer geometry: The
  white (black) sites are the A(B) sub-lattice on which spins transform as the
  fundamental (conjugate) representation of SU($N$). $J_1$ connects nearest neighbors in
  the plane, $J_2$ connect next nearest
  neighbors in the plane and $J_\perp$ connect sites on
  different layers. (b,c) shows cartoon product wavefunctions of
local singlets for the VBL and VBS states. In reality, the
  ground state is a strongly
  interacting superposition of all valence-bond coverings. The ground state 
  nevertheless (b) preserves all symmetries for
  the VBL, but (c) breaks lattice-symmetry (as shown) for the VBS. In
  this paper we provide a detailed study of the N\'eel-VBL and
  N\'eel-VBS quantum phase transitions.}
\end{figure}

In this work we address the destruction of the
SU($N$) symmetry breaking N\'eel
 order in the two-dimensional bilayer system shown in
 Fig~\ref{fig:intro}(a). In the bilayer geometry the Berry phases cancel between the two layers in the continuum limit 
allowing access to the phase transitions of interest without the additional
complication of quantum interference effects. We have studied the
properties of the
phase transitions from N\'eel order to two different types of
paramagnetic states, the valence bond liquid (VBL) and the valence
bond solid (VBS) [see Fig.~\ref{fig:intro}(b,c)]. 
The {\bf N\'eel-VBL} transition for $N=2$ has been studied extensively~\cite{hida1992:bil,millis1993:bil,sandvik1994:bil,wang2006:bilayer} and
is well known to be continuous in the O(3)
 universality class. Here we address for the first time the fate of this
 transition when $N>2$. We find that a simple
 Landau mean-field theory predicts a discontinuous N\'eel-VBL
 transition for $N>2$ and a continuous transitions for $N=2$. Using
 unbiased quantum Monte Carlo simulations we confirm the expectations
 of the Landau theory, except for $N=3$, where we find no evidence for a
 first-order transition. We show that if this transition is
 continuous, its universality class should be identified with a critical point in the compact
 CP$^2$ model~\cite{nahum2011:loops,foot:first}.
The {\bf N\'eel-VBS} transition in the single-layer model has been predicted~\cite{senthil2004:science} and numerically found to be
continuous and in the universality class of the {\em non-compact}
CP$^{N-1}$ model for all $N$~\cite{sandvik2007:deconf,lou2009:sun,kaul2011:j1j2}. We show that remarkably the
N\'eel-VBS transition characterized by the same broken symmetries becomes
first-order  in the bilayer geometry for all $N$ studied here (our
model gives us access to $N\geq 5$), a striking consequence of the
cancellation of Berry phases between layers.
 
{\em Bilayer Model.--}
Our SU($N$) symmetric model is defined with a local Hilbert space of $N$ states on 
each site of the bilayer square lattice illustrated in
Fig.~\ref{fig:intro}(a). 
We label these states as $|\alpha\rangle$
with $1\leq\alpha\leq N$. 
We adopt the representation used previously
in both analytic~\cite{affleck1985:lgN,read1990:vbs,read1989:vbs} and numerical \cite{harada2003:sun,beach2009:sun,kaul2011:j1j2} works on bipartite lattices,
where the sublattice-A states transform under rotations with the
fundamental representation of SU($N$) [generated by the $N^2-1$
matrices $T^a$], and the B sublattice 
states transform with the conjugate of this representation. We
consider two different SU($N$) invariant interaction: between sites $i$
and $j$ on the same sub-lattice $\Pi_{ij} \equiv \sum_a T_i^a \cdot T_j^a$, and
between sites on opposite sub-lattices $P_{ij}\equiv\sum_a T_i^a\cdot T_j^{*a}$.
Using these interactions, we define a model
 SU($N$) symmetric bilayer system as follows,
\begin{equation}
\label{eq:hbil}
H_{\rm bil} = -\frac{J_1}{N} \sum_{\langle ij\rangle} P_{ij} -\frac{J_2}{N} \sum_{\langle\langle ij\rangle\rangle}  \Pi_{ij}-\frac{J_\perp}{N} \sum_{[ ij]}  P_{ij},
\end{equation}
where $\langle ij\rangle$ denotes nearest neighbors in the square
lattice layers,  $\langle\langle ij\rangle\rangle$ denotes
next-nearest neighbors in the square lattice layers and $[ij]$ denotes
inter-layer bonds, as illustrated in Fig.~\ref{fig:intro}(a). The
$J_1$ term by itself gives the familiar single layer SU($N$)
anti-ferromagnet, which is N\'eel ordered for $N\leq 4$ and VBS
ordered for $N\geq 5$. Adding a $J_2$ term to the $J_1$-model favors
the N\'eel state, causing
the N\'eel-VBS transition to move to arbitrary large $N$ as $J_2$ is increased~\cite{kaul2011:j1j2}. Finally, when the $J_\perp$
term is made large enough it always favors the formation of a VBL, by forcing
the formation of local singlets [see Fig.~\ref{fig:intro}(b)].
The model
bilayer anti-ferromagnet, Eq.~(\ref{eq:hbil}), reduces to the 
familiar SU(2) bilayer model 
for $N=2$ and $J_2=0$.

\begin{figure}[t]
\includegraphics[width=3.3in, clip=true, trim= 400 30 0  0]{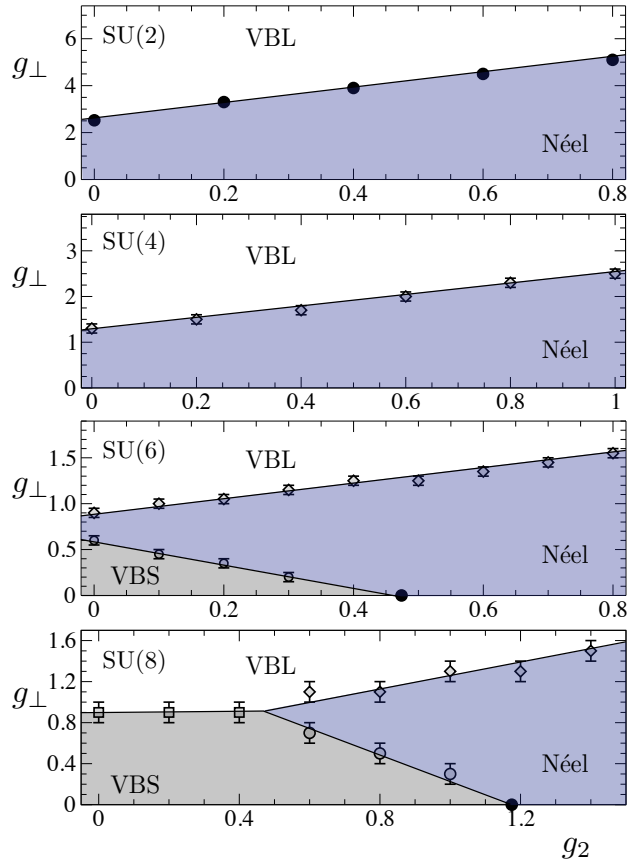}
\caption{\label{fig:pd_su6_su8} Phase diagram of the model $H_{\rm
    bil}$ defined in
  Eq.~(\ref{eq:hbil}) for SU(2), SU(4), SU(6) and SU(8) symmetry in
  the plane of $g_2\equiv J_2/J_1$ and $g_\perp\equiv J_\perp/J_1$. The unfilled symbols are locations of
  first order phase transitions, N\'eel-VBL (diamonds), N\'eel-VBS
  (circles) and VBS-VBL (squares). The solid black circles mark
  continuous transitions. For SU(2), the line of N\'eel-VBL critical points shown
  are in the universality class of the O(3) non-linear $\sigma-$model.
  For SU(6) and SU(8) the N\'eel-VBS transitions shown are in the
  universality class of the non-compact CP$^{N-1}$ models (with
  $N=6,8$ respectively).  Solid lines and shaded regions are guides to the eye.}
\end{figure}

Since $H_{\rm bil}$ satisfies Marshall's sign
criteria, it can be simulated using unbiased quantum Monte Carlo
methods on large lattices of linear dimension $L$ with  $2\times L\times L$ sites and at
finite-temperature $T$ using
the stochastic series expansion method with loop updates~\cite{sandvik1999:sse,evertz2003:loop,sandvik2010:vietri}.
N\'eel order is detected by the existence of a non-zero spin stiffness
$\rho_s= T\langle W^2\rangle$ in the limit of $L\rightarrow\infty$,
where $W$ is the spatial winding number of the world lines~\cite{sandvik2010:vietri}.
Likewise, long-range order in the correlation function $N^2 C_V({\bf r},\tau)=\langle P_{\bf 0,0+x}(0) P_{\bf r,r+x}(\tau) \rangle-\langle P_{\bf 0,0+x}(0) \rangle^2$ signals spontaneous
translational symmetry breaking, {\em i.e.}, the onset of VBS
order. All the VBS ordering studied in our bilayer system is of
the columnar type (at momentum $(\pi,0)$) and is in phase between the
layers (see Fig.~\ref{fig:intro}(c)). We define $O^2_{\rm VBS}$ in the
usual way as the long distance limit of the VBS correlation function. 
Finally, an absence of both long-range N\'eel and VBS orders indicates the
formation of a VBL state.
Using these tests for the three phases, N\'eel, VBS and VBL,  we have computed the $T=0$ phase
diagram in the $g_\perp-g_2$ plane ($g_\perp \equiv J_\perp/J_1$,
$g_2\equiv J_2/J_1$) for each $N\leq 10$. For $N\leq 4$, the model
Eq.~(\ref{eq:hbil}) has only two phases: N\'eel and VBL [Fig.~\ref{fig:intro}(b)]. For $N\geq 5$,
the model admits in addition a VBS phase [Fig.~\ref{fig:intro}(c)]. Phase
diagrams for the bilayer model, Eq.~(\ref{eq:hbil}), for SU(2), SU(4),
SU(6) and
SU(8) symmetry are shown in Fig.~\ref{fig:pd_su6_su8}. These four
cases, contain all the types of phase diagrams we have encountered
in our study with $N\leq 10$. We now turn to the main focus of our paper,
a detailed analysis of the  nature of the N\'eel-VBL and N\'eel-VBS 
phase transitions that appear in these phase diagrams for each $N$.

\begin{figure}[t]
\includegraphics[width=3.5in,trim= 0 0 0 150]{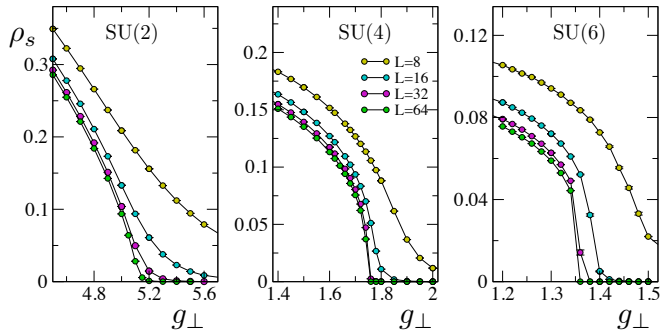}
\caption{\label{fig:first_order} {\bf N\'eel-VBL:} The spin stiffness $\rho_s$ close to
  the N\'eel-VBL transition for SU(2), SU(4) and SU(6).
  The SU(2) transition is continuous and in the O(3) universality
  class. 
The quantity $\rho_s$ for SU(4) and SU(6) show signs of
step-like behavior. Close to the step we find double peaked histograms (see
Fig.~\ref{fig:hist}) characteristic of a
  first-order transition. 
The N\'eel-VBL transition shows such first order
   behavior for all $N\geq 4$.
The parameters used are $g_2=0.8$ for SU(2), $g_2=0.4$
  for SU(4) and  $g_2=0.6$ for SU(6). The legend shows
  the value of $L$, we have set $J_1\beta=L$ everywhere. }
\end{figure}

\begin{figure}[b]
\includegraphics[width=3.4in,clip=true,trim= 0 0 0 320]{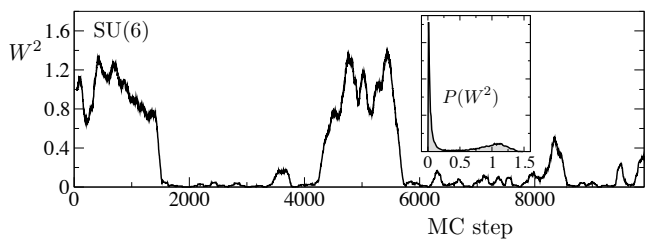}
\caption{\label{fig:hist} {\bf N\'eel-VBL:} Hysteresis and double peaked histograms at a
first order N\'eel-VBL transition in the SU(6) bilayer. In the main
frame we show a sample MC history of the binned squared spatial
winding number, $W^2$, which shows clear evidence for metastability
The inset shows a histogram for the same quantity, with clear double
peaked structure. This
behavior is found only very close to the transition and for
sufficiently large volumes, providing unambiguous evidence for a first
order transition. Here shown for $L=32$, $g_2=0.6$ and $g_\perp=1.36$.}
\end{figure}

{\em N\'eel-VBL.--}
First, we analyze the transition between the N\'eel state and the
featureless fully symmetric
valence bond liquid [a cartoon of the VBL state is illustrated in
Fig.~\ref{fig:intro}(b)]. 
The N\'eel-VBL transition in the bilayer model for $N=2$ and $J_2=0$ has been studied
extensively~\cite{hida1992:bil,millis1993:bil,sandvik1994:bil,wang2006:bilayer}. In the special case of $N=2$ the order parameter
describing the SU(2) symmetry breaking can be
written as an O(3) vector. The absence of Berry phases in the bilayer
geometry then allows for
the description of the critical point in terms of the well known O(3) non-linear
$\sigma-$model~\cite{haldane1988:berry}. This simple mapping has no
known generalization for $N>2$. For general $N$, the simplest description of the N\'eel-VBL phase transition is
found by writing a Landau theory for the order parameter of the
SU($N$) anti-ferromagnet. Such a description contains both the N\'eel
and VBL phases, since the VBL is
featureless and can be thought of simply as a phase in which the SU($N$)
order parameter is quantum disordered. The appropriate order parameter
is an $N\times N$ traceless Hermitian matrix,
${Q}_{\alpha\beta}$, which transforms as $Q\rightarrow U Q U^\dagger$
under SU($N$) rotation. In our model, Eq.~(\ref{eq:hbil}), such a matrix
can be constructed microscopically from a local operator defined as, $\hat Q_{\alpha\beta}(i) \equiv |\alpha
\rangle_i \langle \beta |_i - 1/N$ on the A sub-lattice and $\hat Q_{\alpha\beta}(i) \equiv |\beta
\rangle_i \langle \alpha |_i - 1/N$ on the B sub-lattice. We can now
coarse grain this local operator to obtain the order parameter, $\mathbb{Q}$,  and write down a Landau theory action,
which being SU($N$) invariant must consist of traces of powers of
$\mathbb{Q}$.
\begin{equation}
S_L= \alpha_L {\rm Tr}(\mathbb{Q}^2)+\beta_L {\rm Tr}(\mathbb{Q}^3)+\gamma_L
{\rm Tr}(\mathbb{Q}^4)
\end{equation}
Once the order parameter
acquires an expectation value we can do an SU(N) rotation to obtain a 
 diagonal form for $Q_{\alpha\beta}= m(\delta_{\alpha 1}\delta_{\beta 1}-\delta_{\alpha\beta}/N) $
which is the
analog of a ``collinear'' magnet and the quantity $m$ is the condensate. 
If we now substitute the diagonal form for $\mathbb{Q}$ in $S_L$ we
can see that generally cubic terms in $m$ are present in the
action for $N>2$. In mean-field approach for $N>2$ such 
terms will render the phase transition first order, 
very much like the first order nematic-isotropic transition in liquid crystals~\cite{chaikin2000:cmp}.
 When  $N=2$, it is easy to see that ${\rm Tr} ({\mathbb{Q}^3})$ evaluates to zero and does not give rise
to a cubic $m$ term, making a continuous transition possible. Indeed by
identifying, $n_x=(Q_{12}+Q_{21})/2$,  $n_y=(Q_{12}-Q_{21})/2i$, $n_z=Q_{11}$ and
including gradient terms in the action we
arrive at the 
well known O(3) $\sigma-$model for the $\vec{n}=(n_x,n_y,n_z)$ order
parameter.

\begin{figure}[b]
\includegraphics[width=3.3in,clip=true,trim= 0 0 0 180]{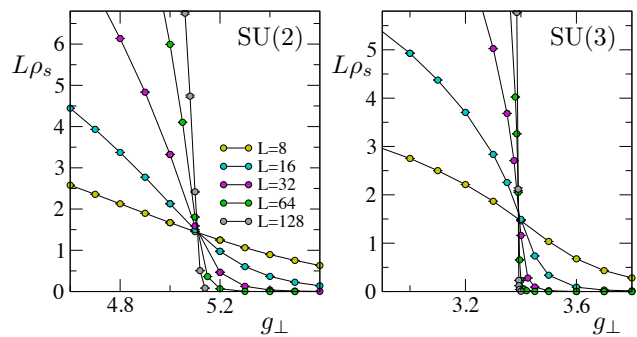}
\caption{\label{fig:su23} {\bf N\'eel-VBL:} Crossings of the fluctuations of the spatial
  winding number at the N\'eel-VBL transition for SU(2) and SU(3). In
  both cases up to sizes of $L=128$ we see good evidence for a nice
  crossing indicating a continuous transition. No evidence for
  first-order behavior was found in these two cases.}
\end{figure}

Consistent with the above Landau theory we confirm from our
numerical simulations (see Figs.~\ref{fig:first_order},~\ref{fig:hist}
and~\ref{fig:su23}) that the
N\'eel-VBL phase transition is continuous for $N=2$ (and in the O(3)
universality class) and first order for $N\geq 4$. The first order
transitions get progressively weaker as $N$ is lowered. Indeed for
$N=3$ we find no evidence for a discontinuous transition up to
$L\leq 128$~\cite{foot:first} (see Fig.~\ref{fig:su23}). If the SU(3) N\'eel-VBL
transition is continuous, what is the
continuum field theoretic description? Does the field theory admit a
critical fixed point? 
The continuum description of the N\'eel-VBL phase
transition in our SU($N$) bilayer Hamiltonian is a 
CP$^{N-1}$ field theory with a {\em compact} U(1) gauge field. In
order to make this connection, we  introduce $N$ complex numbers $z_\alpha$
with the constraint $\sum_\alpha |z_\alpha|^2 = 1$~\cite{read1989:vbs}
and use them to rewrite $Q_{\alpha\beta}=z_\alpha^* z_\beta-\delta_{\alpha\beta}/N$. This
representation has a well known U(1) gauge redundancy which can be
made explicit with the introduction of a gauge field $a_\mu$ in the
long wavelength effective action, the
famous CP$^{N-1}$ model description,
\begin{equation}
\label{eq:cpn}
S = \int d^2x d\tau \left [ \frac{1}{g} |(\partial_\mu-i a_\mu)z_\alpha |^2 +
  F_{\alpha\beta}F_{\alpha\beta}\right ]
\end{equation}
where $F_{\alpha\beta} = \partial_\alpha a_\beta-\partial_\beta
a_\alpha$ is the EM tensor. Following previous work on quantum anti-ferromagnets~\cite{motrunich2004:hhog,senthil2004:deconf_long},
it is clear that in order for the above field theory to possess the VBL
state of the bilayer system when $J_\perp\gg J_1,J_2$, the gauge field $a_\mu$ must be
{\em compact}. The Higgs phase with $ z_\alpha$ condensed corresponds
to a phase with SU($N$) symmetry breaking and we identify this phase
with the N\'eel phase. 
On the other hand, in the phase where $z_\alpha$ is
massive, the photon mode gets confined because of the compactness of
the gauge field and Polyakov's mechanism of monopole proliferation, resulting in a simple fully
gapped paramagnet, which we identify with the VBL phase,
Fig.~\ref{fig:intro}(b). Thus the SU($N$) N\'eel-VBL transition in our
bilayer can be described in the continuum limit by the Higgs-``confined phase''
transition in the {\em compact} CP$^{N-1}$ theory. Recent work~\cite{nahum2011:loops} has found that a lattice discretization of the
compact CP$^{N-1}$ field theory
has a continuous transition for $N=2,3$ and a first order transition
for $N\geq 4$.  Remarkably, this is in full agreement with our findings here
for the SU($N$) bilayer,
strengthening the evidence for our identification of a continuous
transition between N\'eel and VBL for
$N=3$. A detailed study of critical singularities of the SU(3)
N\'eel-VBL fixed point will be presented elsewhere.



\begin{figure}[t]
\includegraphics[width=3.0in,trim= 50 0 0 50]{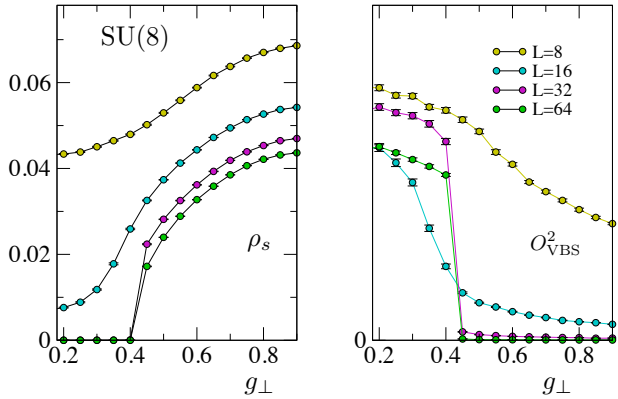}
\caption{\label{fig:nv_first} {\bf N\'eel-VBS:} First order nature of the N\'eel-VBS
  transition in the two dimensional square lattice bilayer. Both
  $O^2_{\rm VBS}$ and $\rho_s$ show evidence for step like behavior at
  the same $g_\perp$. Close to the jump we find the same kind of double peaked
  behavior in $\rho_s$ that is illustrated in Fig.~\ref{fig:hist}. Here we have shown sample data for $N=8$
and $g_2=0.8$. Similar behavior is found for all $N$ studied here.}
\end{figure}

{\em N\'eel-VBS.--} We now turn to the transition between the N\'eel
and translational symmetry breaking valence-bond solid state [the VBS
state is illustrated in Fig.~\ref{fig:intro}(c)]. For a single layer the N\'eel-VBS  transition in
the model defined by Eq.~(\ref{eq:hbil}) was found to be continuous~\cite{kaul2011:j1j2} as
predicted by the ``deconfined'' field theoretic
arguments~\cite{senthil2004:science}. While it is clear that the
N\'eel and VBS phases are individually stable to a small but finite
$g_\perp$, the interlayer coupling is expected to be strongly relevant at
the fixed
point of decoupled deconfined quantum critical points. 
What is the fate of the N\'eel-VBS
transition in the bilayer geometry? From a theoretical point of view,
in the bilayer geometry the cancellation of Berry phases negates the
quantum interference effects that are crucial to the deconfined
quantum criticality scenario~\cite{senthil2004:science}. In the
absence of such effects one expects the restoration of the
conventional Landau paradigm, where the direct transition between two
symmetry breaking states is necessarily first-order independent of the
value of $N$. Indeed as
illustrated in Fig.~\ref{fig:nv_first} from our
QMC simulations we find that the N\'eel-VBS 
phase transition is always first-order in the bilayer geometry. In our
model we only have access to this transition for $N\geq 5$ and in
these cases we always find a first-order transition.
This is a remarkable effect since the phase transition in the
single layer and in the bilayer is in both cases between the
same two phases, {\em i.e.} characterized by exactly the {\em same sets} of
broken symmetries and in the same spatial dimension. The difference in
the long-distance physics between the bilayer and single layer,
much like the Haldane gap in one-dimension, is purely due to the presence/cancellation of
the Berry phases in the single/bi-layer systems.

In conclusion we have presented a detailed analysis of two sets of quantum phase
transitions in bilayer SU($N$) spin systens: 
First, we have studied the fate of the popular~\cite{hida1992:bil,millis1993:bil,sandvik1994:bil,wang2006:bilayer} bilayer SU($N=2$)
{\bf N\'eel-VBL} transition for the case $N>2$, and second, we have studied the fate of the
SU($N$) {\bf N\'eel-VBS}
deconfined critical point~\cite{senthil2004:science,kaul2011:j1j2} for
a single layer in the bilayer geometry. We have found that the $N=2$ continuous N\'eel-VBL
phase transition remains continuous for $N=3$ (in the universality
class of the compact CP$^2$ model~\cite{nahum2011:loops}), becoming first order for $N\geq
4$, and that the cancellation of Berry phases in the bilayer geometry
restores Landau's paradigm for the N\'eel-VBS transition, resulting in a first order phase
transition between two phases with distinct broken symmetries.  

The author gratefully acknowledges useful discussions with L.~Balents,
J.~Chalker, K.~Damle, M.~Levin, A.~Nahum,
A.~Sandvik and A.~Vishwanath. Partial financial
support was received through NSF DMR-1056536. 
The numerical simulations reported in the manuscript were carried out
on the DLX cluster at the University of Kentucky.  


\begin{thebibliography}{00}
\bibitem{foot:first}
Our assertion of a continuous transition for the $N=3$ N\'eel-VBL transition is based on evidence on linear sizes
up to $L\leq 128$. It is clearly not possible to rule out a very weak
first order transition at which $\xi\gg 128$.
\end{thebibliography}

\null

\end{document}